\documentclass[twocolumn,twocolappendix]{aastex702}

\usepackage{amsmath}
\usepackage{graphicx}
\usepackage{threeparttable}
\usepackage{txfonts}
\bibpunct[; ]{(}{)}{;}{a}{}{,}
\hypersetup{allcolors=blue}

\begin{document}
\title{Pushing Variability Searches to Extreme Dwarf Galaxies with the Vera C. Rubin Observatory Data Preview 2}

\author[0000-0002-4436-6923]{Fan Zou}
\email[show]{fanzou01@gmail.com}
\affiliation{Department of Astronomy, University of Michigan, 1085 S University, Ann Arbor, MI 48109, USA}

\author[0000-0002-0167-2453]{W. N. Brandt}
\email{wnbrandt@gmail.com}
\affiliation{Department of Astronomy and Astrophysics, 525 Davey Lab, The Pennsylvania State University, University Park, PA 16802, USA}
\affiliation{Institute for Gravitation and the Cosmos, The Pennsylvania State University, University Park, PA 16802, USA}
\affiliation{Department of Physics, 104 Davey Laboratory, The Pennsylvania State University, University Park, PA 16802, USA}

\author[0000-0001-5802-6041]{Elena Gallo}
\email{egallo@umich.edu}
\affiliation{Department of Astronomy, University of Michigan, 1085 S University, Ann Arbor, MI 48109, USA}

\author[0000-0002-8686-8737]{Franz E. Bauer}
\email{franz.e.bauer@gmail.com}
\affiliation{Instituto de Alta Investigaci\'on, Universidad de Tarapac\'a, Casilla 7D, Arica, Chile}

\author[0000-0002-0101-6624]{Angela Bongiorno}
\email{angela.bongiorno@inaf.it}
\affiliation{INAF-Osservatorio Astronomico di Roma, Via Frascati 33, 00078 Monteporzio Catone, Italy}

\author[0000-0001-9947-6911]{Colin J. Burke}
\email{Colin.Burke@unt.edu}
\affiliation{Department of Physics, University of North Texas, Denton, TX 76203, USA}

\author[0000-0002-6496-9414]{Alister W. Graham}
\email{alister.w.graham@gmail.com}
\affiliation{Centre for Astrophysics and Supercomputing, Swinburne University of Technology, Hawthorn, VIC 3122, Australia}

\author[0000-0001-8416-7059]{Hengxiao Guo}
\email{hengxiaoguo@gmail.com}
\affiliation{Shanghai Astronomical Observatory, Chinese Academy of Sciences, 80 Nandan Road, Shanghai 200030, People's Republic of China}

\author[0000-0002-8606-6961]{Lorena Hern\'andez-Garc\'ia}
\email{lorena.hernandez@mail.udp.cl}
\affiliation{Instituto de Estudios Astrof\'isicos, Facultad de Ingenier\'ia y Ciencias, Universidad Diego Portales, Av. Ej\'ercito Libertador 441, Santiago, Chile}
\affiliation{Centro Interdisciplinario de Data Science, Facultad de Ingenier\'ia y Ciencias, Universidad Diego Portales, Av. Ej\'ercito Libertador 441, Santiago, Chile}

\author[0000-0002-1380-1785]{Marcin Marculewicz}
\email{marcin.marculewicz@gmail.com}
\affiliation{Department of Physics and Astronomy, Wayne State University, 666 W. Hancock St, Detroit, MI 48201, USA}

\author[0000-0002-5678-1008]{Dieu D. Nguyen}
\email{dieun@umich.edu}
\affiliation{Department of Astronomy, University of Michigan, 1085 S University, Ann Arbor, MI 48109, USA}

\author[0000-0002-5854-7426]{Swayamtrupta Panda}\thanks{Gemini Science Fellow}
\email{swayamtrupta.panda@noirlab.edu}
\affiliation{International Gemini Observatory/NSF NOIRLab, Casilla 603, La Serena, Chile}

\author[0000-0003-2196-3298]{Alessandro Peca}
\email{peca.alessandro@gmail.com}
\affiliation{Instituto de Estudios Astrof\'isicos, Facultad de Ingenier\'ia y Ciencias, Universidad Diego Portales, Av. Ej\'ercito Libertador 441, Santiago, Chile}

\author[0000-0001-5711-084X]{Bindu Rani}
\email{binduphysics@gmail.com}
\affiliation{NASA Goddard Space Flight Center, Greenbelt, MD 20771, USA}

\author[0000-0001-5231-2645]{Claudio Ricci}
\email{Claudio.ricci.astro@gmail.com}
\affiliation{Department of Astronomy, University of Geneva, ch. d'Ecogia 16, 1290, Versoix, Switzerland}
\affiliation{Instituto de Estudios Astrof\'isicos, Facultad de Ingenier\'ia y Ciencias, Universidad Diego Portales, Av. Ej\'ercito Libertador 441, Santiago, Chile}
\affiliation{Kavli Institute for Astronomy and Astrophysics, Peking University, Beijing 100871, People's Republic of China}

\author[0000-0003-4327-1460]{Ohad Shemmer}
\email{Ohad.Shemmer@unt.edu}
\affiliation{Department of Physics, University of North Texas, Denton, TX 76203, USA}

\author[0000-0003-0747-1780]{Wei Leong Tee}
\email{wmt5159@psu.edu}
\affiliation{Department of Astronomy and Astrophysics, 525 Davey Lab, The Pennsylvania State University, University Park, PA 16802, USA}
\affiliation{Institute for Gravitation and the Cosmos, The Pennsylvania State University, University Park, PA 16802, USA}

\author[0000-0002-6990-9058]{Zhibo Yu}
\affiliation{Department of Astronomy and Astrophysics, 525 Davey Lab, The Pennsylvania State University, University Park, PA 16802, USA}
\affiliation{Institute for Gravitation and the Cosmos, The Pennsylvania State University, University Park, PA 16802, USA}
\email{zvy5225@psu.edu}

\begin{abstract}
Galaxies with stellar masses $M_\star<10^8~M_\odot$, hereafter referred to as extreme dwarf galaxies (EDGs), represent one of the least explored regimes in the search for accreting intermediate-mass black holes. Using the recent Vera C. Rubin Observatory Data Preview 2 (DP2; covering 2025-04-25 to 2026-01-06), we present the first systematic search for variability among EDGs in the ELAIS-S1 field, the most extensively observed LSST Deep-Drilling Field in DP2. Applying single-band and cross-band variability statistics to difference-imaging light curves for $6.2\times10^4$ EDGs with archival multiwavelength characterizations, we identify seven variable EDG candidates, all brighter than $i$-band magnitude of 24 and located within $z<0.3$. The two brightest candidates are previously known AGNs, three have light curves more consistent with supernovae, and two do not have a clear nature. The sample corresponds to an apparent EDG variable fraction of $6.8_{-2.1}^{+3.1}\times10^{-4}$ at $i$-band magnitudes brighter than 24 and is expected to consist primarily of AGNs and supernovae. Compared to a magnitude-matched massive-galaxy sample with the same variability selection applied, the EDG variable fraction is comparable or at most lower by up to a factor of three. This proof-of-concept study demonstrates that Rubin will identify a statistically meaningful sample of variable EDG candidates, opening a new regime for black-hole demographics. Spectroscopic follow-up will nevertheless be required to secure redshifts and confirm individual candidates.
\end{abstract}


\section{Introduction}
\label{sec: intro}
Over the past decade, extensive efforts have been made to extend our understanding of supermassive black holes (SMBHs) and their massive galaxy hosts (e.g., \citealt{Kormendy13}) to lower-mass galaxies and their likely intermediate-mass black holes (IMBHs; e.g., \citealt{Greene20}), through active galactic nuclei (AGNs; e.g., \citealt{Baldassare18, Mezcua18, Davis21, Burke22, Ward22, Salehirad22, Zou23, Wasleske24, Pucha25}), tidal disruption events (TDEs; e.g., \citealt{Jin25, Wang26}), and dynamical searches (e.g., \citealt{Nguyen19, Davis20, Zaw20, Haberle24}). 
In the coming years, the Legacy Survey of Space and Time (LSST; \citealt{Ivezic19}) of the Vera C. Rubin Observatory (Rubin) is expected to revolutionize IMBH searches by monitoring their time-domain signals. Its unprecedented depth, cadence, and sky coverage will yield much tighter constraints on IMBHs than previous surveys, and extend variability-based searches to many more, and previously unexplored, galaxies at the low-mass end (e.g., \citealt{Burke23, Burke26}).\par

Although thousands of AGNs have been identified in dwarf galaxies with stellar masses $M_\star<3\times10^9~M_\odot$,\footnote{Some studies have also adopted a threshold of $M_\star<10^{10}~M_\odot$ to define ``low-mass galaxies''. In any case, these definitions are much more massive than our EDGs.} most of these samples are still concentrated at the relatively high-mass end ($M_\star\sim10^9~M_\odot$). We still have almost \textit{zero} knowledge of black holes at $M_\star\lesssim10^8~M_\odot$ due to observational challenges of confirming their AGNs. We hereafter refer to galaxies with $M_\star<10^8~M_\odot$ as extreme dwarf galaxies (EDGs) and focus on this more extreme regime in the context of AGN studies and IMBH candidate searches. In this work, we use the Rubin Data Preview 2 (DP2; \citealt{RTN115}) to conduct the first systematic variability characterization of EDGs.\par

Black holes in EDGs represent an important, fundamentally different regime with possibly unanticipated phenomena (e.g., \citealt{LSST09, Norris17, Li22}). The $M_\star$ gap between EDGs and the Large Magellanic Cloud (LMC; \citealt{vanderMarel02}) is as large as that between the LMC and the Andromeda Galaxy \citep{Tamm12}. Such orders-of-magnitude mass differences would result in qualitatively different environments for their resident black holes, including much shallower potential wells, lower metallicities, and more stochastic star-formation histories (e.g., \citealt{Tolstoy09, Annibali22, Sales22}). The extremely low $M_\star$ would also indicate that their resident black holes are highly likely to be IMBHs, although the $M_\mathrm{BH}-M_\star$ scaling relation is largely unconstrained at $M_\star\lesssim10^9~M_\odot$. The need to push down to the EDG regime has also been revealed by population-level analyses of more massive dwarf galaxies. For example, the black-hole occupation fraction remains statistically uncertain in EDGs (e.g., \citealt{Zou25}), yet this regime is generally more sensitive to black-hole seeding and growth models. At $M_\star\gtrsim10^8~M_\odot$, the observed $M_\mathrm{BH}$ may also be largely built up through cosmic accretion (F. Zou et al. submitted), and thus signatures of the initial black-hole seeds may be much cleaner in EDGs than in more massive dwarf galaxies.\par

However, black holes in EDGs have been nearly inaccessible to previous facilities. As W.~L.~Tee et al. (submitted) showed, current \mbox{X-ray} facilities are not sufficiently sensitive over large sky areas to discover a statistically meaningful EDG AGN population, although a few \mbox{X-ray} candidates have been detected in EDGs within $\sim$50~Mpc, whose nature remains debated \citep{Zou25}. AGN diagnostics based on emission-line ratios also largely fail at low metallicities \citep{Cann19}. Therefore, optical variability is one of the most promising methods for black-hole searches in EDGs. Previous variability-selected dwarf AGN samples have almost bottomed out near $10^9~M_\odot$ due to shallower sensitivities (e.g., \citealt{Baldassare18, Baldassare20, Burke22, Ward22}). However, Rubin can probe magnitudes deeper than previous time-domain facilities, such as the Zwicky Transient Facility (ZTF; \citealt{Bellm19}), enabling variability studies of EDGs. As simulated in \citet{Burke23}, LSST is expected to identify variable AGNs well below $M_\star=10^8~M_\odot$.\par

In this work, we present a proof-of-concept study to examine whether variable EDGs can be detected in the latest Rubin DP2, the first data release obtained with the LSST Science Camera (LSSTCam). Since IMBH AGNs are predicted to show optical variability dominated by timescales of days to weeks \citep{Burke21}, a few months of Rubin DP2 data are better matched to IMBH studies than to SMBHs. As a pilot study, this work focuses on the most extensively observed of the planned LSST Deep-Drilling Fields (DDFs), the European Large Area ISO Survey S1 (ELAIS-S1) field, with a nominal center at R.A. = 00:38:00 and decl. = $-44$:00:00. We show that optical variability in EDGs is already detectable with a few months of Rubin observations, with the sample expected to comprise both AGNs and supernovae (SNe). With a limited baseline, transient contamination is the defining challenge in this regime, and we quantify it against independent rate predictions.\par

We adopt a flat $\Lambda\mathrm{CDM}$ cosmology with $H_0=70~\mathrm{km~s^{-1}~Mpc^{-1}}$, $\Omega_\Lambda=0.7$, and $\Omega_M=0.3$. Throughout this work, we quote IDs from the ELAIS-S1 catalog of \citet{Zou22} rather than the cataloged object IDs from Rubin DP2.

\section{Data and Parent Sample}
\label{sec: data}
Rubin conducted a series of engineering tests, on-sky commissioning, and early observations before the formal start of LSST. DP2 comprises LSSTCam observations obtained between 2025-04-25 and 2026-01-06, including early on-sky engineering observations, the main Science Validation survey, and part of the pre-LSST observations between the construction-to-operations handover. It covers $\approx15000~\mathrm{deg^2}$ in total, of which $\approx3000~\mathrm{deg^2}$ were cataloged, including several deep fields. We refer readers to \citet{RTN115} for further technical details about DP2.\par

Within the Rubin footprint, the five DDFs will receive more extensive LSST observations than the wide survey, and their central parts have carefully compiled multiwavelength data (e.g., \citealt{Brandt18, Zou22, Weaver22}). Among the DDFs, ELAIS-S1 received the most extensive Rubin observations in DP2 and surpasses the other DDFs in both depth and light-curve baseline.\footnote{The COSMOS field, another DDF, reaches a similar depth to ELAIS-S1, but it does not have comparable baseline.} Therefore, we focus on the ELAIS-S1 field in this work to search for variable EDG candidates using DP2. It has $\approx770$ Rubin visits over 62 nights between 2025-06-30 and 2026-01-06, including 57, 153, 128, 232, 141, and 59 visits across 10, 30, 23, 39, 23, and 11 nights in the $u$, $g$, $r$, $i$, $z$, and $y$ bands, respectively.\par

The parent galaxy sample is drawn from the ELAIS-S1 catalog of \citet{Zou22}, which includes 0.8 million objects over the central $3.4~\mathrm{deg^2}$ and compiles archival multiwavelength data and provides source properties measured through spectral energy distribution (SED) fitting. The archival optical-to-IR photometry was drawn from the multi-band forced-photometry catalog of \citet{Zou21a}, and the photometric redshifts (photo-$z$s) were derived by fitting the $u$-band-to-$4.5~\mu\mathrm{m}$ SEDs in \citet{Zou21b}. We first select $7\times10^4$ EDGs with $M_\star<10^8~M_\odot$ and $z>0.05$. The latter criterion drops $\approx30\%$ of the EDGs in \citet{Zou22} and reduces the effects of unreliable photo-$z$s, most of which pile up at $z=0.01$ (as noted in \citealt{Zou21b}), as well as those of peculiar motions in the very local universe.\par

We match these sources to DP2 through the \texttt{LSDB} package \citep{Caplar25} with a $1''$ radius; false matches will not affect our results because all candidates are visually inspected in Section~\ref{sec: selection}. We retain only objects with successful photometry in at least one Rubin band and without pixel-level flags, including saturation, cosmic rays, interpolated pixels, detector edges, missing data, and pixels clipped or
rejected during coaddition, at their centers, leaving $6.2\times10^4$ EDGs. We then extract the difference-imaging light curves from the \texttt{psfDiffFlux} and \texttt{psfDiffFluxErr\_corrected} columns in \texttt{LSDB}, where the latter provides corrected uncertainties such that non-variable objects have reduced $\chi^2$ close to unity (K. Malanchev et al. in preparation).

\section{Selection of Candidate Variable Extreme Dwarf Galaxies}
\label{sec: selection}
Following the general approaches of \citet{Burke22} and \citet{Ward22}, we nonparametrically select variable objects as those that show both single-band variability and cross-band variability. We will also empirically calibrate the variability selection using all the ELAIS-S1 objects in \citet{Zou22}. Since we are focusing on a single field, these objects shared the same observing conditions and strategies, enabling a meaningful self-calibration.\par
For the first criterion, we follow procedures similar to those of \citet{Shen19} and \citet{Burke22}, but with the following statistical modifications. For a single-band light curve with photometric measurements $f_i$ and uncertainties $\sigma_i$, we have
\begin{align}
f_i\sim\mathcal{N}(\mu, \sigma_0^2+\sigma_i^2),\label{eq: f_whitenoise}
\end{align}
where $\mu$ and $\sigma_0^2$ are free parameters. The log-likelihood is thus
\begin{align}
\ln\mathcal{L}=-\frac{1}{2}\sum_i\left[\ln(\sigma_0^2+\sigma_i^2)+\frac{(f_i-\mu)^2}{\sigma_0^2+\sigma_i^2}+\ln(2\pi)\right].
\end{align}
By solving $\frac{\partial\ln\mathcal{L}}{\partial\mu}=0$, we obtain the profile likelihood as a function of $\sigma_0^2$:
\begin{align}
\ln\mathcal{L}_p(\sigma_0^2)=\ln\mathcal{L}\left(\mu=\frac{\sum_i\frac{f_i}{\sigma_0^2+\sigma_i^2}}{\sum_i\frac{1}{\sigma_0^2+\sigma_i^2}}\right).\label{eq: lnLp}
\end{align}
To test against the constant-flux model ($\sigma_0^2=0$), the corresponding maximum likelihood ratio is
\begin{align}
\Delta\ln\mathcal{L}_\mathrm{max}=\max_{\sigma_0^2\ge0}\{\ln\mathcal{L}_p(\sigma_0^2)\}-\ln\mathcal{L}_p(0).
\end{align}
This is then converted to a Gaussian-equivalent significance level (e.g., \citealt{Cowan11}):
\begin{align}
Z_\mathrm{single}=\sqrt{2\Delta\ln\mathcal{L}_\mathrm{max}}.
\end{align}\par
For two light curves in bands $a$ and $b$, we conduct a test similar to that of \citet{Welch93}, but again with the following statistical modifications. This tests both variability and cross-band correlation and, as discussed in \citet{Welch93}, is robust against spurious outlying data points occurring in only one band (see also Appendix~\ref{append: clean}). For this method, we bin the light curves so that observations in the two bands form photometric pairs $(f_{a,i}, f_{b,i})$. We then define the paired residuals as
\begin{align}
x_i=\frac{f_{a,i}-\overline{f_a}}{\sigma_{a,i}},\\
y_i=\frac{f_{b,i}-\overline{f_b}}{\sigma_{b,i}},
\end{align}
where $\overline{f_a}$ and $\overline{f_b}$ are the weighted means of the corresponding light curves. If there is no variability and $x_i$ and $y_i$ are uncorrelated, we expect
\begin{align}
Z_\mathrm{cross}=\frac{\sum_i{w_ix_iy_i}}{\sqrt{\sum_iw_i^2}}\sim\mathcal{N}(0, 1),
\end{align}
where $w_i$ denotes a set of predefined weights. Mathematically, it can be shown that the optimal weight for signal detection is
\begin{align}
w_i=\frac{1}{\sigma_{a,i}\sigma_{b,i}},
\end{align}
which we adopt in this work. $Z_\mathrm{cross}$ thus defines the cross-band variability significance.\par
We apply the above variability metrics to the Rubin difference-imaging light curves. We first clean all the Rubin detector-visits because a small fraction of them can contribute to spurious variability, as discussed in detail in Appendix~\ref{append: clean}. We then bin the raw light curves nightly (typically 5 visits for each band) and focus on internight variability, because intranight variability is generally difficult to probe reliably and is more sensitive to systematics (e.g., \citealt{Zuo26}). The characteristic optical variability timescales\footnote{This timescale refers to the single-band scale. Two-band lags are usually much shorter (e.g., even down to minutes; \citealt{Pan26}), which, however, is irrelevant here for single-band variability. The short lag also indicates that it has negligible impacts on $Z_\mathrm{cross}$.} of IMBHs and SMBHs are expected to exceed one day, whereas only stellar-mass compact objects below $\approx10^2~M_\odot$ vary on timescales below 0.5~days \citep{Burke21}. Nightly binning therefore should not materially affect the detection of IMBHs and SMBHs, while suppressing other contamination. The intranight variability with Rubin, however, is still scientifically important and should be probed in future dedicated work. We further apply $3\sigma$ clipping (e.g., \citealt{Burke22, Wasleske22}) to the $f/\sigma$ distribution of each light curve. A small fraction of light curves contain single-visit outliers, either physical or instrumental. Since AGN light curves rarely show such sudden changes, clipping suppresses these outliers while leaving AGNs largely unaffected, although some fast transients might be excluded.\par

$Z_\mathrm{single}$ is then calculated from the nightly binned light curves for each band. For $Z_\mathrm{cross}$, we instead further bin the light curves in three-night intervals, since Rubin observed ELAIS-S1 in $griz$ every $\approx2-3$ nights. We find that three-night binning maximizes the number of photometric pairs, with a median of 14--23 for any two-band combination of $griz$. We have verified that our selection is generally robust to the choice of this bin width.\par

We regard an object as variable if the maximum (over different bands) $Z_\mathrm{single}>5$ and $Z_\mathrm{cross}>3$. The slightly lower threshold for $Z_\mathrm{cross}$ reflects its role as the second criterion: applied only to objects that already satisfy $Z_\mathrm{single}>5$, it need not carry the same false-positive burden. It also has intrinsically lower power, since cross-band pairing retains only $\approx50\%-80\%$ of the nights available to $Z_\mathrm{single}$. This overall selection returns 20 variable candidates. We then inspect their Rubin images and find that 13 of them appear likely to be contaminated by nearby bright sources. We compare their Rubin photometry with the archival photometry in \citet{Zou22} and find that the Rubin measurements of many objects significantly deviate from the overall SED baseline, and their Rubin photometry may be biased by nearby sources during standard pipeline processing. Therefore, we retain only the remaining 7 isolated objects in our final sample. The median ratio between their $\sigma_0$ and total fluxes is 0.1.\par

\begin{figure*}
\includegraphics[width=\linewidth]{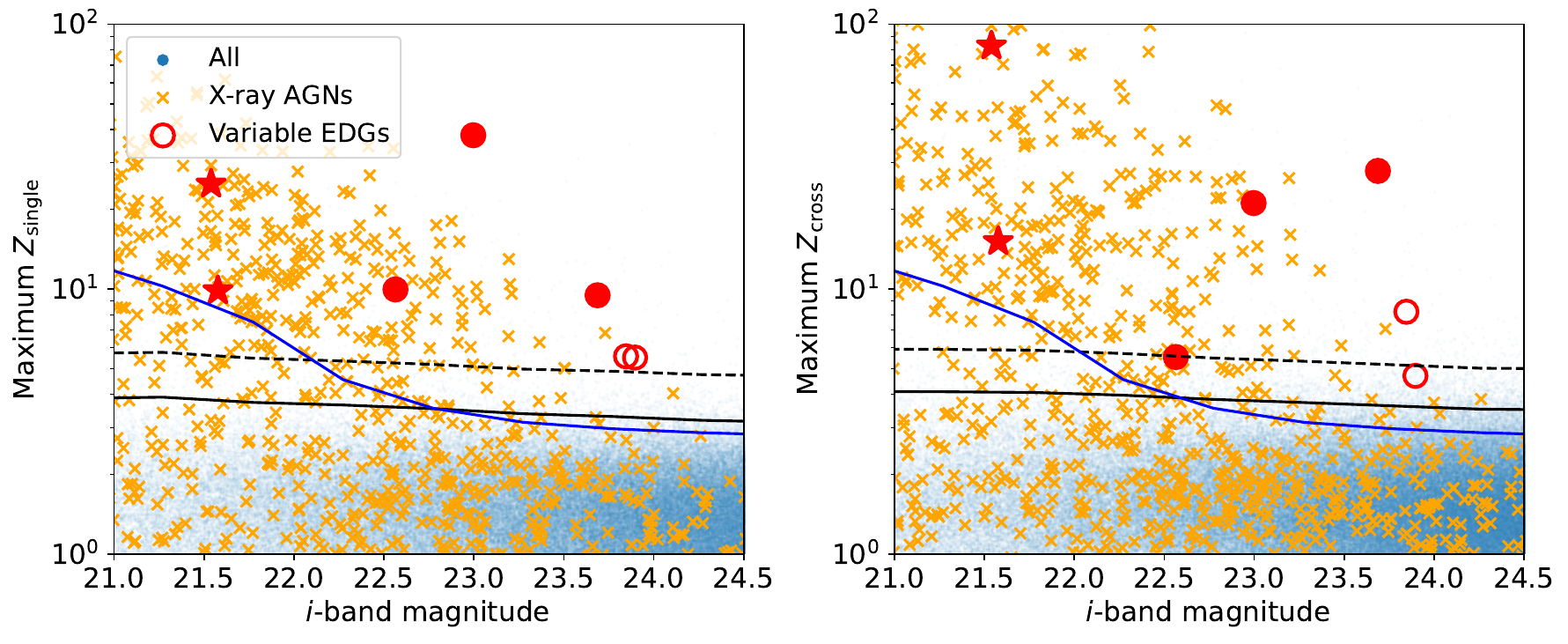}
\caption{The maximum $Z_\mathrm{single}$ (left) and $Z_\mathrm{cross}$ (right) versus Rubin $i$-band magnitude. The red symbols are our identified variable EDG candidates, where the two red stars, three filled circles, and two open circles represent the flagged AGNs in \citet{Zou22} (Section~\ref{subsec: agn_signal}), the SNe-like objects, and those without clear nature, respectively. The orange crosses are \mbox{X-ray} AGNs in \citet{Ni21}. The background blue points represent all the ELAIS-S1 objects in \citet{Zou22}, based on which the expected variability thresholds corresponding to $3\times$ and $5\times$ population dispersion are derived and plotted as the black solid and dashed curves, respectively. The blue curves represent the $99^\mathrm{th}$ percentiles of all the ELAIS-S1 objects.}
\label{fig: Zvsmag}
\end{figure*}

Figure~\ref{fig: Zvsmag} shows the maximum $Z_\mathrm{single}$ and $Z_\mathrm{cross}$ versus $i$-band magnitude. The selected variable EDG candidates reach $i$-band magnitudes down to 24~mag. For comparison and self-calibration, we derive empirical thresholds in each magnitude bin as median + $X\times$ ($84^\mathrm{th}$ percentile $-$ median), with $X=3$ and 5, from all ELAIS-S1 objects in \citet{Zou22}, assuming that most objects are not variable. We also show the $99^\mathrm{th}$ percentile curves of maximum $Z_\mathrm{single}$ and $Z_\mathrm{cross}$ in Figure~\ref{fig: Zvsmag}, which decrease toward fainter magnitudes, consistent with the expectation that, at fixed $Z$ thresholds, the apparent variability fraction would decrease at faint magnitudes (see Appendix~\ref{append: clean}). We have verified that the curves remain nearly identical when using only EDGs for calibration. All selected variable EDG candidates lie well above the $5\times$ dispersion curve in $Z_\mathrm{single}$, the $3\times$ curve in $Z_\mathrm{cross}$, and the extreme 1\% tail curves. For comparison, the \mbox{X-ray}-selected AGNs in \citet{Ni21} are explicitly labeled in Figure~\ref{fig: Zvsmag}, and most of those brighter than $i\approx22.5$ are recovered as variable. The variability fraction of these \mbox{X-ray}-selected AGNs decreases toward fainter magnitudes, possibly in part because they are more likely to be obscured, although detailed analyses of these massive AGNs are beyond the scope of this work. Our variable EDG candidates are generally fainter than the \mbox{X-ray}-selected AGNs, illustrating Rubin's unique ability to open such a low-mass regime, which was nearly inaccessible to previous facilities.\par

With several procedures applied, including detector-visit cleaning, light-curve clipping, the inclusion of $Z_\mathrm{cross}$, and self-calibration using objects in the same field, we have taken steps to assess and mitigate potential sources of spurious variability. Visual inspection of their light curves does not reveal any obvious issues, and most objects indeed show apparent variability patterns (Section~\ref{subsec: lc}). We have verified that the variability selection of most objects is insensitive to the single most significant nights, with the only exception of 644246932909, which has relatively small $Z_\mathrm{single}$ and $Z_\mathrm{cross}$ among the sample.\par
\begin{figure*}
\includegraphics[width=\linewidth]{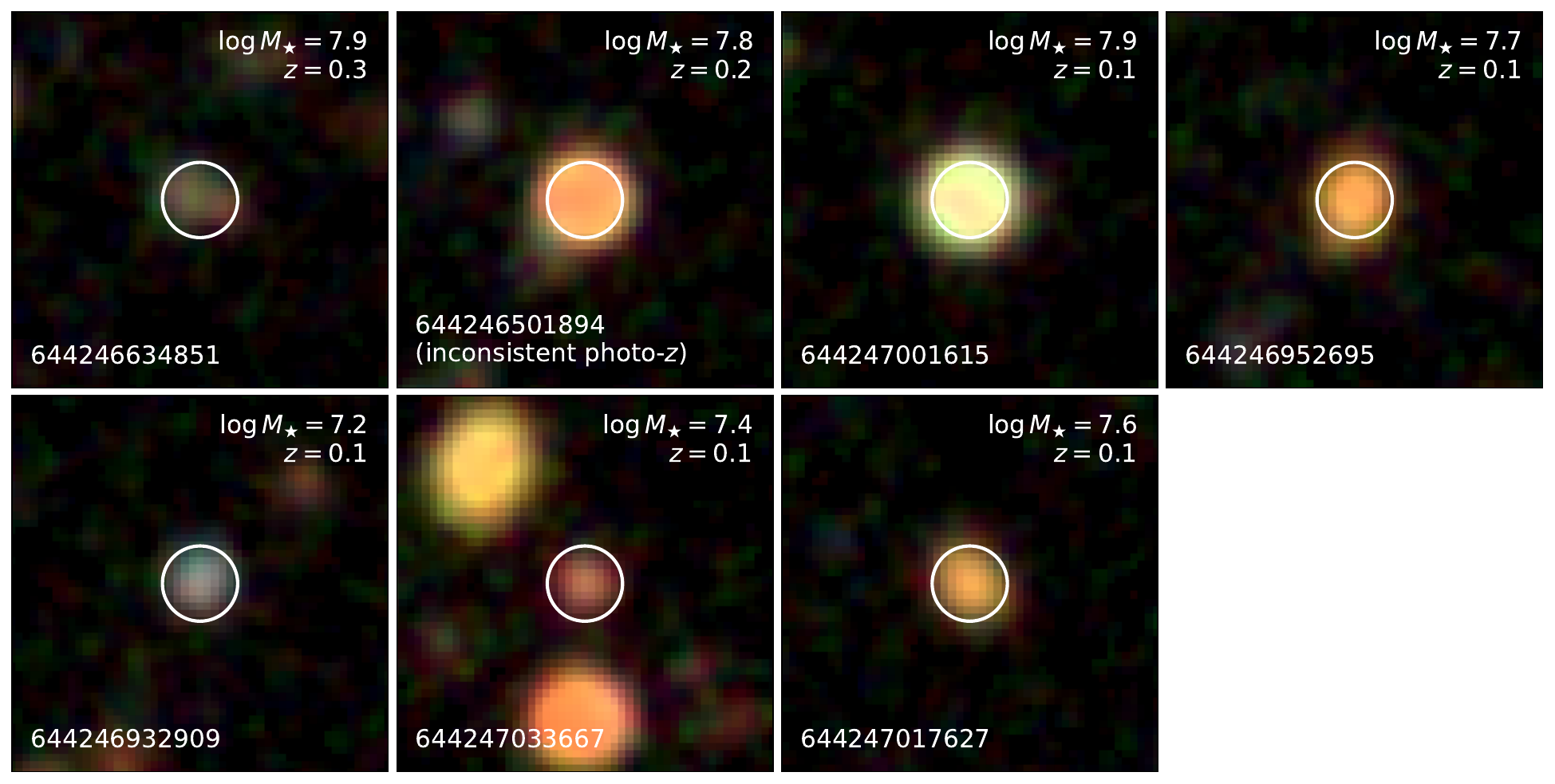}
\caption{Rubin $10''\times10''$ three-color cutouts of the final sample, generated from $g$-band (blue), $r$-band (green), and $i$-band (red) images and smoothed for visualization purposes. The white circles are centered on the targets and have radii of $1''$. Their object IDs from \citet{Zou22}, $M_\star$, and redshifts are labeled in each panel. The object with a likely unreliable photo-$z$ in Appendix~\ref{append: zphot} is explicitly marked.}
\label{fig: imgs_all}
\end{figure*}
\begin{figure}
\includegraphics[width=\linewidth]{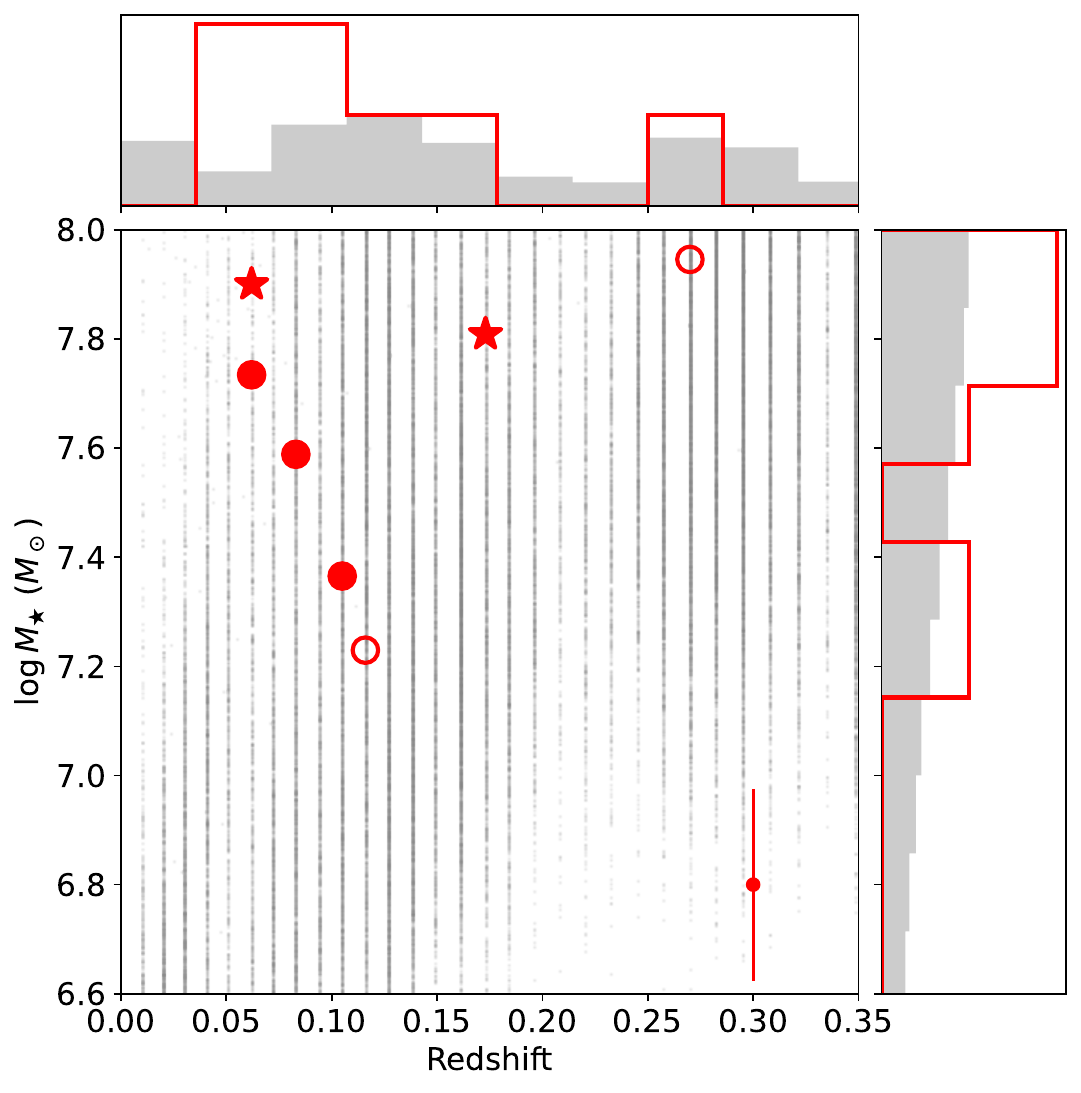}
\caption{The identified variable EDG candidates (red symbols) in the $M_\star$-redshift plane, plotted in the same manner as in Figure~\ref{fig: Zvsmag}. Their typical $1\sigma$ $M_\star$ uncertainties are plotted at the bottom-right corner. The background gray points are all the galaxies in ELAIS-S1, and their redshifts appear quantized because of the chosen photo-$z$ grid in \citet{Zou21b}. The top and right panels show the normalized histograms of $\log M_\star$ and $z$ of the variable EDG candidates (red) and all the galaxies (gray).}
\label{fig: Mz}
\end{figure}

We plot the Rubin $gri$ three-color images in Figure~\ref{fig: imgs_all}. Most of them are faint and relatively compact, as expected for EDGs. We further plot the $M_\star$ and redshifts of our final sample in Figure~\ref{fig: Mz}, and all the objects are within $z<0.3$. Compared to the parent EDG galaxy sample, the variable EDG candidates tend to reside in the regime with higher $M_\star$ and lower redshifts. They are generally well detected in the near-infrared \citep{Jarvis13} and thus have at most moderate median $M_\star$ statistical uncertainties in \citet{Zou22} of 0.2~dex, which, however, do not include redshift uncertainties. We instead expect that the dominant uncertainty would be from their photo-$z$s. None of our selected objects has a spec-$z$, and photo-$z$s and distances may be unreliable for individual objects, as assessed in Appendix~\ref{append: zphot}. We therefore regard them only as candidates, and spectroscopic follow-up will be required to secure redshifts.

\section{Nature of the Variable EDG Candidates}
\label{sec: nature}
\subsection{Known AGNs}
\label{subsec: agn_signal}

The two brightest objects (IDs 644246501894 and 644247001615) in our sample of seven variable EDG candidates were previously flagged as AGNs based on multiwavelength data, and they are also \mbox{X-ray}-selected AGNs in \citet{Ni21} with XIDs ES0359 and ES1993. Their SEDs in \citet{Zou22} are not fully AGN-dominated for $M_\star$ estimate purposes. Specifically, their AGN-model $\log M_\star$ values are 7.8 and 7.9 for IDs 644246501894 and 644247001615, respectively; when all emission is attributed to the galaxies, the corresponding galaxy-model $\log M_\star$ values are 8.3 and 7.9, which represent soft upper limits on the actual $M_\star$. Therefore, their main uncertainties remain their redshifts, as discussed in Appendix~\ref{append: zphot}. These objects demonstrate that Rubin variability selection can identify genuine AGNs, although spectroscopic follow-up observations remain necessary to confirm individual candidates.

\subsection{Light-curve Modeling}
\label{subsec: lc}

We analyze the Rubin light curves of the selected sample in this section. Equation~\ref{eq: f_whitenoise} is a white-noise model. We further consider two phenomenological models: (i) a generic burst model similar to that of \citet{vanVelzen19}, which is suitable for transients, and (ii) a damped random walk (DRW) model suitable for AGNs (e.g., \citealt{Kelly09}).\par

For the former, we fit only the $g$- and $r$-band difference-flux light curves to allow for direct comparisons with previous works (e.g., \citealt{vanVelzen19, vanVelzen21}). The $g$-band light curve $f_g(t)$ is modeled as
\begin{align}
f_g(t)=\begin{cases}
f_\mathrm{peak}e^{-\frac{(t-t_\mathrm{peak})^2}{2\sigma^2}}+c_g, t\le t_\mathrm{peak}\\
f_\mathrm{peak}e^{-\frac{t-t_\mathrm{peak}}{\tau}}+c_g, t>t_\mathrm{peak}
\end{cases}
\end{align}
where $f_\mathrm{peak}$ is the peak flux, $t_\mathrm{peak}$ is the peak time, $\sigma$ is the rising timescale, $\tau$ is the fading timescale, and $c_g$ is a constant. During this early phase of Rubin operations, the sky templates may contain part of the transient emission; we therefore include a free constant term $c_g$ to account for possible offsets from the pre-transient sky. The $r$-band light curve is modeled simultaneously with the $g$-band light curve using a color term:
\begin{align}
(g-r)(t)&=\begin{cases}
\langle g-r\rangle, t\le t_\mathrm{peak}\\
\langle g-r\rangle+a(t-t_\mathrm{peak}), t>t_\mathrm{peak}
\end{cases}\\
f_r(t)&=[f_g(t)-c_g]10^{0.4(g-r)(t)}+c_r,
\end{align}
where $\langle g-r\rangle$ is the pre-peak color (assumed to be constant) of the transient event, $a$ is the color-change rate, and $c_r$ accounts for any template offset due to possible transient contamination.\par

When this model is applied to SNe, TDEs, and SMBH AGNs, these populations usually have different fitted parameters. Specifically, SNe usually have relatively small, week-scale $\sigma$ and $\tau$; TDEs have slightly larger, month-scale $\sigma$ and $\tau$; and SMBH AGNs mainly populate the longest, year-length timescales. However, IMBH AGNs are expected to vary on shorter timescales and thus may overlap with SNe and TDEs. SNe tend to show significant color-change rates, with a large fraction having $a\gtrsim0.01~\mathrm{day^{-1}}$, while TDEs and AGNs generally have near-zero color-change rates, with rare exceptions such as changing-state AGNs (e.g., \citealt{Ricci23}). Therefore, the burst-modeling results may provide first-order clues about the nature of the light curves.\par

Similarly, the DRW model provides a phenomenological description of the correlated stochastic variability observed in AGNs. Its variability timescale is characterized by $\tau_\mathrm{DRW}$, which has been found to correlate with the mass of the central compact object \citep{Burke21}. Based on the scaling relation in \citet{Burke21}, the expected rest-frame $\tau_\mathrm{DRW}$ is above 19~days for SMBHs above $10^6~M_\odot$ and decreases to $1.3-19$~days for IMBHs between $10^3-10^6~M_\odot$. However, reliably measuring $\tau_\mathrm{DRW}$ requires a long baseline spanning $\gtrsim10\tau_\mathrm{DRW}$ \citep{Kozlowski17}, and the DRW model may not be sufficiently accurate over this short DP2 timescale, as found in previous AGNs (e.g., \citealt{Zu13, Stone22}). Therefore, the DP2 light curves are still insufficient to characterize SMBH AGN variability.\par

\begin{figure*}
\includegraphics[width=\linewidth]{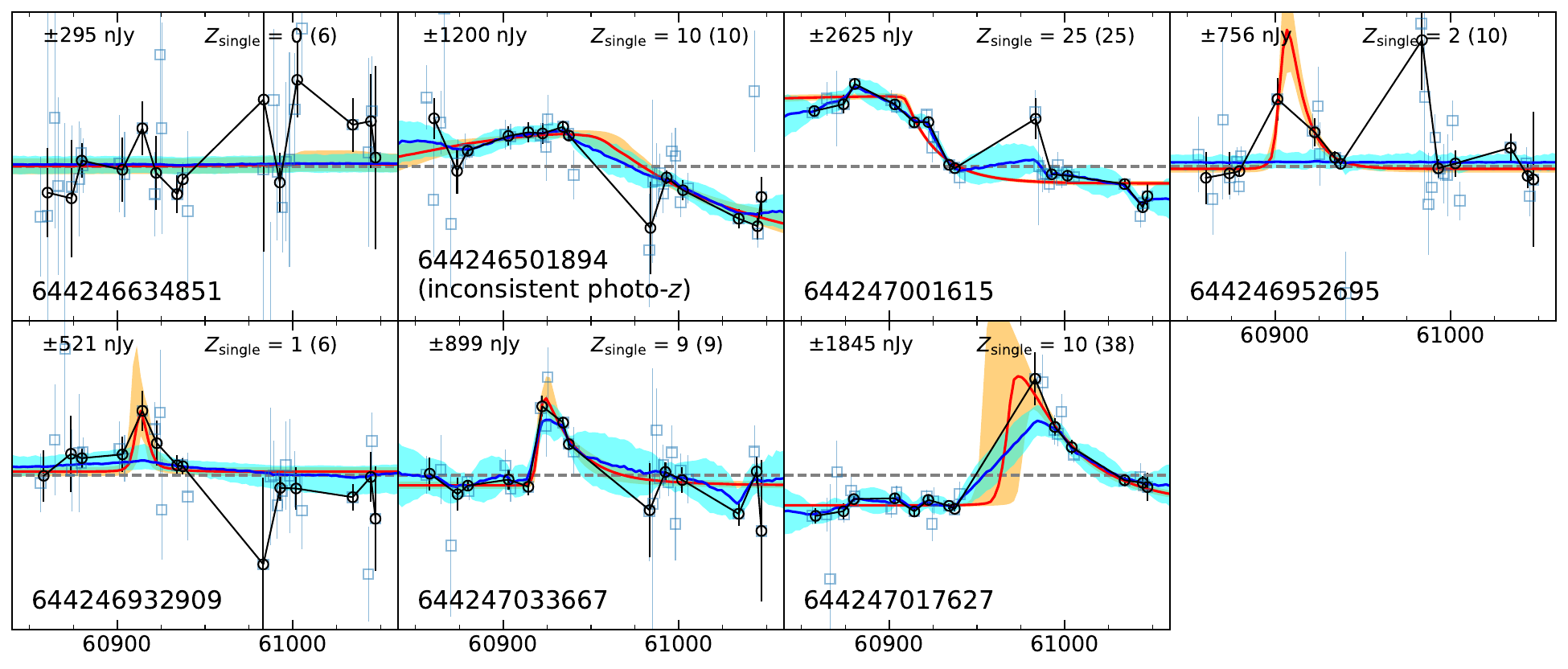}
\caption{The $g$-band difference-flux light curves of the selected sample. The background blue squares with $1\sigma$ error bars are the nightly binned light curves. The black points are binned every 10 days for visualization purposes only. The red and blue curves represent the fitted burst and DRW models and their $1\sigma$ uncertainties, respectively. The horizontal black dashed lines mark zero flux, and the plotted ordinate-axis ranges (fluxes in nano-Jansky) are listed in the top-left corner of each panel. The top-right corner of each panel lists the $g$-band $Z_\mathrm{single}$ and the maximum $Z_\mathrm{single}$ out and in the parentheses, respectively. The bottom-left corners label the IDs in \citet{Zou22}, where the object with a likely unreliable photo-$z$ in Appendix~\ref{append: zphot} is explicitly marked.}
\label{fig: lcfit}
\end{figure*}

We fit the $gr$ light curves with the burst model using \texttt{emcee} \citep{Foreman-Mackey13} and fit each band with the DRW model using \texttt{EzTaoX} \citep{Yu26}, adopting flat priors for all fits. The fitted $g$-band light curves are shown in Figure~\ref{fig: lcfit}. The burst and DRW models generally provide similar quality fits to the data over the limited Rubin DP2 timespan. To quantitatively assess the fits, we compare their Bayesian information criteria (BICs) with that of the white-noise model. The BIC difference is defined as $\Delta\mathrm{BIC}=2\Delta\ln\mathcal{L}_\mathrm{max}-\Delta k\ln N$, where $\Delta k$ is the difference in the number of free parameters, and $N$ is the number of data points. A large $\Delta\mathrm{BIC}\ge6$ is considered strong evidence favoring the more complex model. Considering the $gr$ light curves, 6/7 objects have statistically significant evidence against the white-noise model. When the other bands are considered, the remaining object, ID 644246634851, also statistically disfavors the white-noise model in the $i$ band (not explicitly plotted). Therefore, all of our sample show statistically significant structure beyond a white-noise model.\par

Based on visual inspections of the light curves, three objects appear more transient-like (likely SNe): 644246952695,\footnote{644246952695 is not apparently transient-like in the $g$ band shown in Figure~\ref{fig: lcfit}, but it is more obvious in $r$ and $i$, which are not explicitly plotted.} 644247033667, 644247017627. The Rubin data cover both the rising (with a time scale of a few days) and fading (a few weeks) phases for them in Figure~\ref{fig: lcfit}. Therefore, our sample includes two known AGNs (Section~\ref{subsec: agn_signal}), three SNe-like objects, and two objects without clear nature.\par

\begin{figure*}
\includegraphics[width=\linewidth]{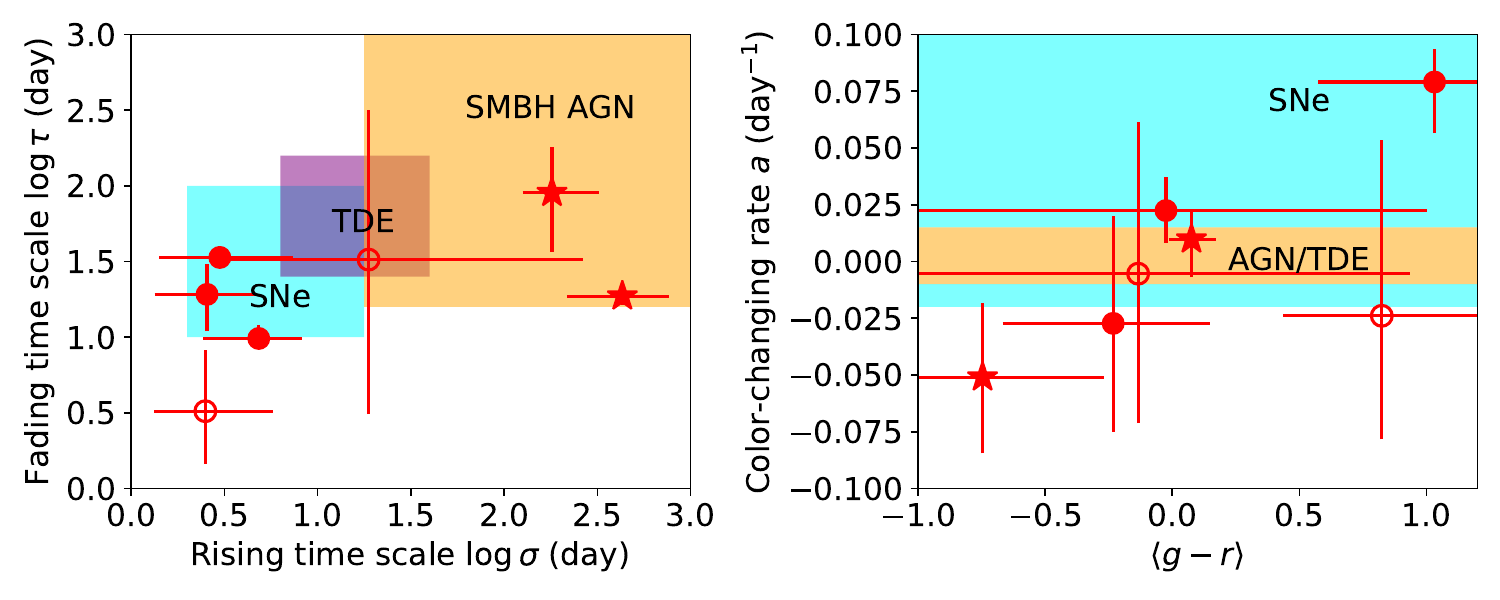}
\caption{The light-curve fitting results for the burst model, where the red symbols are plotted following the same manner as in Figure~\ref{fig: Zvsmag}. The left panel shows the burst timescale $\sigma$ versus $\tau$, and the right panel shows the $\langle g-r\rangle$ versus $a$ plane. The approximate regions populated by different types of objects, adapted from \citet{vanVelzen21}, are shown and labeled.}
\label{fig: lcparams}
\end{figure*}

We further show the fitted burst-model parameters in Figure~\ref{fig: lcparams}. Given the limited number of available observing nights, the fitting is often only indicative and has limited diagnostic power. In the $\sigma-\tau$ diagram, the two known AGNs (Section~\ref{subsec: agn_signal}) fall in the SMBH AGN regime, and the three SNe-like objects reside in the SN locus, while the two objects without clear nature have large uncertainties and do not have apparent clues. We stress that, given the limited timespan of DP2, detailed light-curve classifications are still uncertain, and we only treat our physical labels as exploratory.

\section{Summary}
\label{sec: summary}

This work identifies seven variable EDG candidates using the Rubin DP2 data in ELAIS-S1, the most extensively observed of the planned LSST DDFs in DP2. Detailed fitting of the Rubin light curves indicates that all objects statistically require more complex models than white noise. The two brightest ones were previously classified as AGNs, three objects have SNe-like light curves, and two objects have unknown nature, although the classification is only exploratory because of the limited Rubin data. Overall, this analysis sheds light on an almost entirely unexplored, and very low-mass regime and demonstrates Rubin's unique potential for discovering variable EDGs.\par

The identified candidate count corresponds to an apparent variable fraction of $6.8_{-2.1}^{+3.1}\times10^{-4}$ at $21.5<i<24$ and a sky density of $2~\mathrm{deg^{-2}}$ at $M_\star<10^8~M_\odot$. Based on the expected SN rate from the scaling relations in \citet{Sullivan06} and \citet{Madau14}, the mean SN fraction within the DP2 window is $3.3\times10^{-4}$ for the EDG parent sample at $i<24$. This expectation is consistent with our observed SN rate ($2.9_{-1.3}^{+2.3}\times10^{-4}$), regarding all the SNe-like objects in Section~\ref{subsec: lc} as SNe. Therefore, SNe could account for a substantial fraction of variable EDGs. On the other hand, extrapolating the TDE rate of \citet{Yao23} down to EDGs yields an expected TDE fraction of only $10^{-5}$ within the DP2 window. Therefore, the identified variable EDGs are expected to comprise primarily SNe and AGNs, both at approximately the $10^{-4}$ level. The longer baseline of LSST Data Release 1, the next major release after DP2, should substantially improve the discrimination between these populations.\par

To compare the EDG variable fraction with more massive galaxies, we construct an ELAIS-S1 sample with $M_\star>10^{10}~M_\odot$ and $i$-band magnitudes matched to the EDG magnitude distribution at $i<24$. With the same variability selection applied, the apparent massive-galaxy variable fraction is $2.2\times10^{-3}$, higher than the EDG value by a factor of three. However, since our visual inspection retains only 35\% of the variable EDG candidates (Section~\ref{sec: selection}), EDGs still likely have a variable fraction comparable to that of massive galaxies once the latter is scaled by the same factor.\par

The area probed here is only $3.4~\mathrm{deg^2}$, the central part of ELAIS-S1 with the best archival multiwavelength data \citep{Zou22}. Even a simple extrapolation to the $60~\mathrm{deg^2}$ covered by the DDFs at the current rate suggests that LSST stands to identify hundreds of variable EDGs in future data releases, and likely many more in the wide survey. Indeed, using Rubin DP2 alone, hundreds of variable dwarf galaxies at $M_\star<3\times10^9~M_\odot$ have already been identified in ELAIS-S1 (private communication with W. L. Tee), potentially increasing the number of known dwarf AGNs ($\approx20$; \citealt{Burke22, Zou23}) in ELAIS-S1 by an order of magnitude. Such large samples will enable a transition from individual discoveries to detailed population-level studies and provide exciting opportunities to characterize IMBHs and investigate their possible connections to early black-hole seeds. For example, as simulated in \citet{Burke23}, different black-hole seeding models are expected to be distinguishable with variable EDGs from the full LSST data.\par

Nevertheless, one important caveat concerns the photo-$z$ reliability, which may be limited for individual objects. Future datasets combining Rubin, the Euclid mission, and the Nancy Grace Roman Space Telescope will enable further improvements of photo-$z$s. As Rubin discovers an increasing number of variable EDG candidates, the need for spectroscopic follow-up observations will become increasingly important, and a well-characterized sample of variable EDG candidates will be important for designing follow-up observations.

\begin{acknowledgments}
We thank Jialai Wang for helpful discussions. WNB, WLT, and ZY acknowledge support from NSF grant AST-2407089 and the Penn State Eberly Endowment. FEB acknowledges support from ANID-Chile BASAL CATA FB210003 and FONDECYT Regular 1241005. LHG acknowledges financial support from ANID program FONDECYT Iniciaci\'on 11241477. SP is supported by the International Gemini Observatory, a program of NSF NOIRLab, which is managed by the Association of Universities for Research in Astronomy (AURA) under a cooperative agreement with the U.S. National Science Foundation, on behalf of the Gemini partnership of Argentina, Brazil, Canada, Chile, the Republic of Korea, and the United States of America. AP acknowledges financial support from ANID/FONDECYT/Postdoctorado No. 3260848. CR acknowledges support from SNSF Consolidator grant F01$-$13252 and ANID BASAL project FB210003. This material is based upon work supported in part by the National Science Foundation through Cooperative Agreements AST-1258333 and AST-2241526 and Cooperative Support Agreements AST-1202910 and 2211468 managed by the Association of Universities for Research in Astronomy (AURA), and the Department of Energy under Contract No. DE-AC02-76SF00515 with the SLAC National Accelerator Laboratory managed by Stanford University. Additional Rubin Observatory funding comes from private donations, grants to universities, and in-kind support from LSST-DA Institutional Members. This publication is based in part on proprietary Rubin Observatory Legacy Survey of Space and Time (LSST) data, and was prepared in accordance with the Rubin Observatory data rights and access policies. All authors of this publication meet the requirements for co-authorship of proprietary LSST data. This research uses services or data provided by the Rubin Science Platform at NSF-DOE Vera C. Rubin Observatory, which is jointly funded by the U.S. National Science Foundation and the U.S. Department of Energy, Office of Science.
\end{acknowledgments}

\appendix
\section{Detector-visit Cleaning}
\label{append: clean}
We define a detector-visit as the data from one of the 189 LSSTCam science charge-coupled devices in a single visit, and we found that a small fraction of them can significantly contribute to spurious variability, especially toward the faint end (see the next paragraph). To empirically correct this, we use all the ELAIS-S1 objects within the footprint of \citet{Zou22} with $21<i<25$ and compute their residual signal-to-noise ratios $x=(f-\overline{f})/\sigma$ in each detector-visit ($6.6\times10^4$ in total), where $f$ and $\sigma$ are \texttt{psfDiffFlux} and \texttt{psfDiffFluxErr\_corrected}, respectively. For each detector-visit, we evaluate the mean and standard deviation of $x$ (after clipping to $|x|<5$) and the fraction of objects with $|x|>3$. The majority of the detector-visits have mean $x$ values well close to 0 and standard deviations close to 1, as expected, but a small fraction deviate. To empirically clean these, we remove detector-visits for which any of the three metrics lies more than $3\sigma$ from all detector-visits, which excludes 5\% of detector-visits.\par
\begin{figure}
\includegraphics[width=\linewidth]{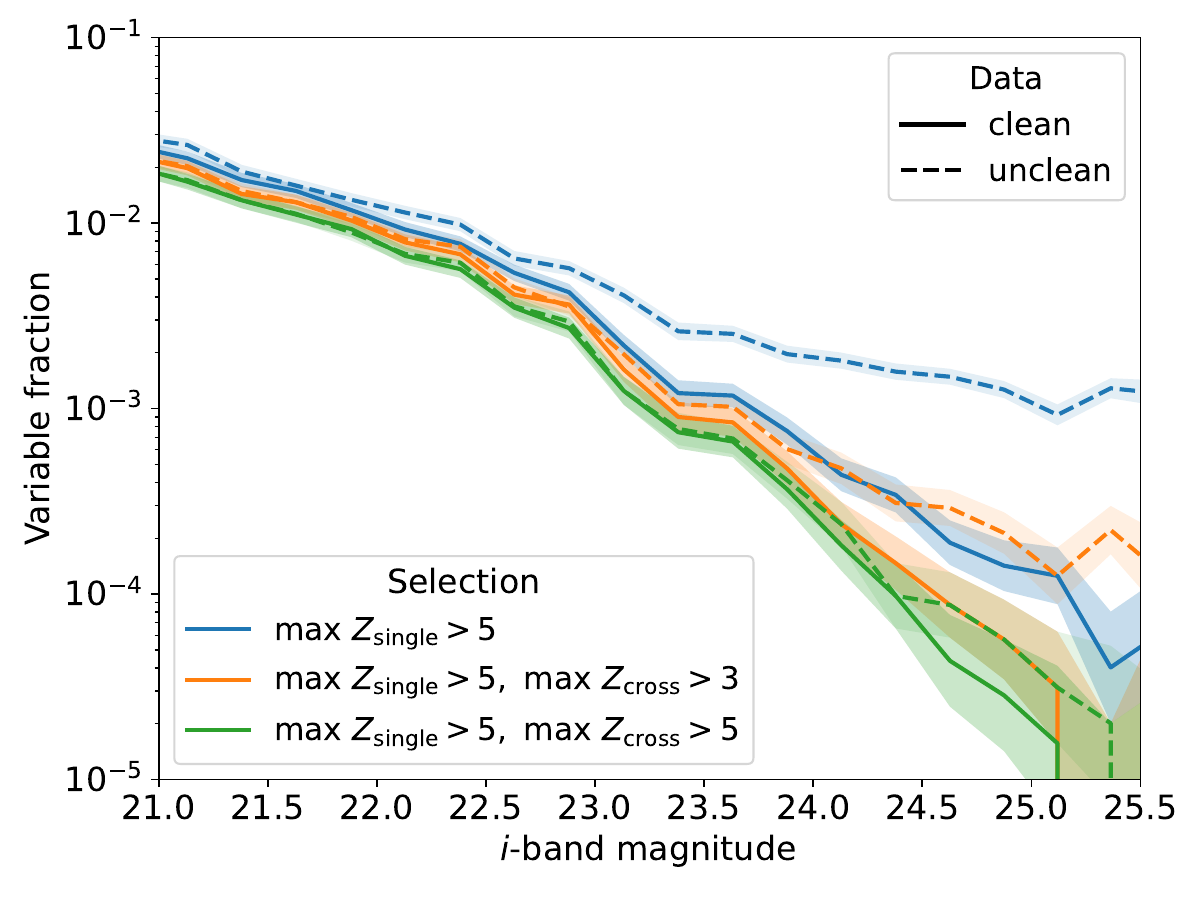}
\caption{The variable fraction versus the $i$-band magnitude with (solid) or without (dashed) detector-visit cleaning under different $Z_\mathrm{cross}$ criteria, as labeled. The shaded regions show the $1\sigma$ binomial confidence intervals. Both the $Z_\mathrm{cross}$ criterion and the cleaning can effectively remove suspiciously variable objects at the faint end.}
\label{fig: fvar_clean}
\end{figure}

The impact of this cleaning is profound for variability selection. Figure~\ref{fig: fvar_clean} compares the variable fractions versus the $i$-band magnitude with and without the detector-visit cleaning under different thresholds of the maximum $Z_\mathrm{cross}$. Without either the detector-visit cleaning or the $Z_\mathrm{cross}$ criterion, the variable fraction tends to be significantly overestimated, especially at the faint end. However, once the cleaning is applied, most of the variable candidates at the faint end would be dropped. On the other hand, the $Z_\mathrm{cross}$ criterion can also effectively clean the same suspiciously variable objects, which usually have low $Z_\mathrm{cross}$. If adopting a sufficiently high threshold ($>5$) for the maximum $Z_\mathrm{cross}$, the detector-visit cleaning would not affect the variable fraction materially. Therefore, the cleaning and the $Z_\mathrm{cross}$ criterion serve as two independent safeguards for reliable variability selection.

\section{Photometric Redshifts}
\label{append: zphot}
At such low stellar masses, none of our variable EDG candidates has an available spec-$z$, so the analysis must rely on photo-$z$s. Due to the limited spectroscopic depth in ELAIS-S1, even the whole EDG parent sample in \citet{Zou22} lacks sufficient spec-$z$s for comparison. However, redshifts and distances are not necessarily accurate for individual objects for at least four reasons: (i) variability may distort the observed SEDs; (ii) the photo-$z$ quality may significantly degrade in the low-mass regime due to, e.g., shallower Balmer breaks; (iii) the photo-$z$ quality degrades toward faint magnitudes; and (iv) the conversions from redshifts to distances have larger uncertainties at low redshifts. This has been one of the fundamental limitations of EDG studies, and spectroscopic follow-up observations will ultimately be necessary. Fortunately, our objects are all brighter than the effective photo-$z$ depth in ELAIS-S1 ($i\approx24$; \citealt{Zou21b}), and the photo-$z$ quality degrades significantly only beyond this depth.\par

Here, the reliability of the photo-$z$s is instead assessed through cross-validation between estimates obtained using different methods. As shown in \citet{Salvato22}, if the photo-$z$s measured with different packages agree, they are much more likely to be closer to spec-$z$s. As a similar illustration using the EDGs within the COSMOS field, we compare their spec-$z$s \citep{Khostovan26} and photo-$z$s \citep{Weaver22} measured with \texttt{EAZY} \citep{Brammer08} and \texttt{LePhare} \citep{Arnouts99, Ilbert06, Shirley26}. Among the objects whose two types of photo-$z$s are inconsistent, i.e., $|z_\mathrm{EAZY}-z_\mathrm{LePhare}|/(1+z_\mathrm{EAZY})>0.15$, $\approx80\%$ are catastrophic outliers with respect to their spec-$z$s, whereas this fraction drops to $\approx10\%$ for those with consistent photo-$z$s. These values are only illustrative and not directly quantitatively applicable to ELAIS-S1 because COSMOS has slightly deeper multiwavelength depths.\par

The adopted photo-$z$s were originally derived by \citet{Zou21b} using \texttt{EAZY}. We then remeasure the photo-$z$s of the selected variable objects using \texttt{LePhare}. The archival photometry in \citet{Zou21a} includes the $u$ band from the VST Optical Imaging of the CDF-S and ELAIS-S1 Fields (VOICE; \citealt{Vaccari16}), $BVR$ from the ESO-Spitzer Imaging extragalactic Survey (ESIS; \citealt{Berta06, Vaccari16}), $grizY$ from the Dark Energy Survey (DES; \citealt{Abbott21}), $ZYJHK_s$ from the VISTA Deep Extragalactic Observations (VIDEO; \citealt{Jarvis13}), and 3.6 and $4.5~\mu\mathrm{m}$ from the Spitzer Survey of Deep-Drilling Fields (DeepDrill; \citealt{Lacy21}). We also supplement these archival data with the deep Rubin $ugrizy$ photometry. For our variable EDG candidates, typically 16 of these bands have good signal-to-noise ratios above five.\par
Among the seven variable objects identified in Section~\ref{sec: selection}, four (57\%) have consistent \texttt{EAZY}- and \texttt{LePhare}-based photo-$z$s according to the classical definition of $|z_\mathrm{EAZY}-z_\mathrm{LePhare}|/(1+z_\mathrm{EAZY})<0.15$. However, 2/3 of the outliers do not appear fully catastrophic but are instead close to the boundary. Under a slightly looser threshold of $|z_\mathrm{EAZY}-z_\mathrm{LePhare}|/(1+z_\mathrm{EAZY})<0.2$, 6 (86\%) objects would have mutually consistent photo-$z$s. The only apparent outlier is 644246501894, one of the two known AGNs, and we thus regard it as more exploratory. Additionally, at low redshifts, distance measurements are sensitive to small redshift changes. If instead adopting $z_\mathrm{LePhare}$, three more objects (644246952695, 644247033667, and 644247017627) would exceed the nominal EDG $M_\star$ threshold of $10^8~M_\odot$ but are still within $10^9~M_\odot$. Therefore, approximately $50\%-80\%$ of our objects appear more reliable, and follow-up spectroscopic observations will still be needed eventually.

\bibliography{citations}{}
\bibliographystyle{aasjournalv7}
\end{document}